# MULTIUSER DETECTION AND CHANNEL ESTIMATION FOR MULTIBEAM SATELLITE COMMUNICATIONS

Helmi Chaouech[1] and Ridha Bouallegue[2]

[1]National Engineering School of Tunis, University of El-Manar, Tunis, Tunisia
`helmi.chaouech@planet.tn`
[1,2]Innov'Com Laboratory, Sup'Com, University of Carthage, Tunis, Tunisia
`ridha.bouallegue@gnet.tn`

*ABSTRACT*

*In this paper, iterative multi-user detection techniques for multi-beam communications are presented. The solutions are based on a successive interference cancellation architecture and a channel decoding to treat the co-channel interference. Beams forming and channels coefficients are estimated and updated iteratively. A developed technique of signals combining allows power improvement of the useful received signal; and then reduction of the bit error rates with low signal to noise ratios. The approach is applied to a synchronous multi-beam satellite link under an additive white Gaussian channel. Evaluation of the techniques is done with computer simulations, where a noised and multi-access environment is considered. The simulations results show the good performance of the proposed solutions.*

## 1. INTRODUCTION

Wireless communications became a required alternative to overcome some deficiencies of wired solutions. They offer mobility and flexibility for users. Thus, data transmission and telephony, such as examples of communications services, are no longer depended on place or movement of the mobile terminal. This behoof of using wireless communications and its several advantages are due to their same wireless support of transmission; the atmosphere which, unlike wired support, is everywhere in every time. However, transmitted signals via atmosphere are subject to many problems such as multi-path, reflection, attenuation, absorption, noise, interference etc. To solve these problems, powerful systems are needed in the basic stations.

Satellite systems, with use of wireless support of transmission, can replace hundreds of thousands of cables and optical fibres. Thus, they can cover and offer communications services to, in addition of urban areas; rural, marine and Saharan zones. The question now is how to access and use a shared wireless support for satellite communications.

Multibeam technology is a good solution for the use of frequency resource. It consists to form at the satellite a throng of narrow beams instead of a single wide beam. In other words, it divides a wide frequency band into some narrow frequency bands. That leads to a regular reuse of the frequency resource and therefore, a significant increasing of the system capacity. Thus, each carrier frequency, which characterises a beam, is a common channel shared between the users of the zone covered by this beam. Thereafter, in order to use the same frequency, users, which are in the same zone or cell, can adopt a TDMA or CDMA multiple access etc.

Our intervention is intended to contribute to solve some problems in this area of communications. In this paper, we have proposed some solutions of multi-user detection and channel estimation in order to deal with interference between co-frequency channels, which is called the co-channel interference, CCI. We developed iterative methods of channel estimation, interference cancellation and channel decoding. These techniques take advantage from the spatial diversity due to antenna array. This operation is realized with extraction of useful signal from all co-frequency or adjacent beams.

The remainder of this paper is organized as follows. Section 2 gives an overview of the related work. Section 3 details the signals models adopted in the multi-user detection and channel estimation techniques. In section 4, we have showed the architecture of the system with its different functional blocks, and detailed with mathematical expressions, the operation of each receiver unit. Section 5 presents the simulations results and gives some analyzes and interpretations of them. Finally, in section 6, we have drawn some conclusions from the proposed solutions and presented some ideas for the future work.

## 2. RELATED WORK

In wireless communications field, channel estimation and detection are subjects of significant importance. That's why, in the last two decades, an important amount of works are developed in these fields, particularly in that of detection techniques. These tasks are generally jointly solved. In fact, detection which consists to extract data from noisy signal needs channel effects compensation. In [1], some channel estimation techniques and propagation delay estimation methods are developed and evaluated. In [2], some wireless channel models are discussed and compared. A solution of channel estimation technique integrated into a multi-user detection algorithm is detailed in [3]. Other solutions of channel estimation can be found in [4], [5], [6] and [7]. Detection algorithms can be evaluated and compared by computation of a bit error rate (BER), and the complexity of their implementation. In [8], Verdu proposed, analyzed and evaluated some detection techniques. Thus, these last can be classified into categories based on their mathematical formulation. The conventional detector or the matched filtering detector suffers from multiple access interference (MAI), and is very sensitive to near-far problem [8], [9]. In [10], the maximum likelihood (ML) detector is developed to solve optimally the deficiencies of the conventional detector. Thus, this ML based detection technique has optimal performance in the presence of MAI and near-far problem at the cost of computational complexity which does not promote its practical implementation. The decorrelator and the minimum mean-squared error (MMSE) detector are linear multi-user detection techniques. They combat MAI and are robust against near-far problem [8], [11], [9]. But, noise enhancement and inversion of big dimension matrices, especially in the case of asynchronous data transmissions, are their basic computation weakness. The interference cancellation multi-user detectors are robust solutions which subtract MAI and improve the BER iteratively [12], [13], [14]. The detection techniques can deal with signals by cancelling MAI serially or with parallel processing. Thus, they can be divided into two main architectures; the successive interference cancellers (SIC) and the parallel interference ones (PIC). To combat noise and interference; channel coding is a good alternative to minimize the BER. Thereafter, multi-user detection solutions, which are based on channel decoding and interference cancellation, show high performance even with low signal to noise ratios (SNR) [15],

International Journal of Computer Networks & Communications (IJCNC) Vol.4, No.1, January 2012

[16]. Some multi-user detection methods, which are developed for multibeam communications, are detailed in [17], [18]. Other detection solutions implementing channel decoding can be found in [19], [20], [21] and [22]. These solutions are based on forward error correction (FEC) offered by channel coding. Thus, detection quality is improved in each decoding iteration of the decoder. Finally, some detection techniques, which cancel the multiple access interference, are presented and evaluated in [23], [24], [25], [26] and [27].

## 3. SIGNALS MODELS

We consider the reverse link of a multibeam satellite system. K active users transmit their signals to the satellite antenna array. The signal of the $k^{th}$ user can be expressed by:

$$r_k(t) = a_k e^{j\varphi_k} x_k(t) \quad (1)$$

Where, $a_k$ and $\varphi_k$ are the amplitude of the signal and its carrier phase, and $x_k(t)$ is given by:

$$x_k(t) = \sum_{i=0}^{N-1} x_k[i] g(t - iT - \tau_k) \quad (2)$$

With, $x_k[i], i = 0..N-1$ is a sequence of N QPSK symbols, $g(t)$ is the emitter filter waveform, $T$ is the symbol temporal duration and $\tau_k$ is the propagation delay of the $k^{th}$ signal. As we treat the synchronous communications, then we take; $\tau_1 = \tau_2 = \ldots = \tau_K = 0$. The sequence of $N$ symbols, which are convolutional coded and interleaved before transmission, is divided into $N_p$ pilot and $N_i$ information symbols. This frame is depicted as follows:

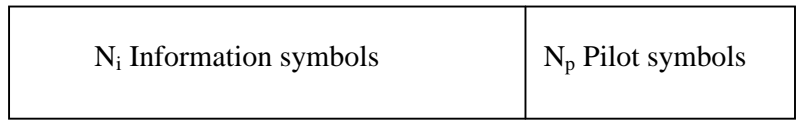

| $N_i$ Information symbols | $N_p$ Pilot symbols |

Figure 1.  Data frame structure.

This signal, expressed in (2), is received by the L radiating components of the antenna array. Then, if we take into account the K users, the signal received by the $l^{th}$ source of the antenna can be expressed by:

$$s_l(t) = \sum_{k=1}^{K} d_k^{(l)} r_k(t) + n_l(t) \quad (3)$$

where, $d_k^{(l)}$ is the $l^{th}$ element of the steering vector $d_k$, of the $k^{th}$ received signal, and $n_l(t)$ is an additive gaussian noise, added to the composite signal received by the $l^{th}$ antenna sensor.
We define the vector comprising the L signals, received by the antenna array sensors, by: $s(t) = [s_1(t), s_2(t), \ldots, s_L(t)]^T$. The vector representation of equation (3), taking account the L components of the antenna, is given by:

$$s(t) = \sum_{k=1}^{K} d_k r_k(t) + n(t) \qquad (4)$$

with, $d_k$ is the column vector of length L which contains information about the arrival direction of the k[th] signal [28], and $n(t) = [n_1(t),\ldots,n_L(t)]^T$ is the additive noise vector at the L sensors outputs. We define the direction of arrival (DOA) matrix of the K beams by : $D = [d_1,\ldots,d_K]$, which is composed of vectors $d_k; k = 1..K$, and $r(t) = [r_1(t),\ldots,r_K(t)]^T$. Where, $[.]^T$ denotes the transpose operator. Thereafter, relation (4) can be rewritten also:

$$s(t) = Dr(t) + n(t) \qquad (5)$$

Beam forming operation maximizes the energy of the useful signal by steering the antenna array in the DOA of this beam [28]. Then, to form the k[th] beam, the treatment consists of taking a linear combination of the signals at the antenna elements outputs. Thus, the k[th] signal is:

$$y_k(t) = v_k^T s(t) \qquad (6)$$

where, $v_k$ is the column vector, which contains the L coefficients of the k[th] beam forming. Generalization of the expression (6) for the K actives users in the system can be expressed as:

$$y(t) = Vs(t) \qquad (7)$$

Where, $y(t) = [y_1(t), y_2(t),\ldots, y_K(t)]^T$ and $V = [v_1, v_2,\ldots,v_K]^T$.

By combining equations (5) and (7), we obtain:

$$y(t) = VDr(t) + Vn(t) = Wr(t) + Vn(t) \qquad (8)$$

We define the diagonal matrix of the of the K signals complex amplitudes by: $A = diag([a_1 e^{j\varphi_1},\ldots,a_K e^{j\varphi_K}])$. With, $diag(.)$ denotes the matrix diagonal operator. Using expression (1), equation (8) can be rewritten as follows:

$$y(t) = WAx(t) + Vn(t) \qquad (9)$$

With, $x(t) = [x_1(t),\ldots,x_K(t)]^T$. By introducing the channel matrix $H = WA$ and the noise vector $z(t) = Vn(t)$, equation (9) becomes:

$$y(t) = Hx(t) + z(t) \qquad (10)$$

After optimal sampling, the model described in (10) can be expressed as:

$$y[i] = Hx[i] + z[i] \qquad (11)$$

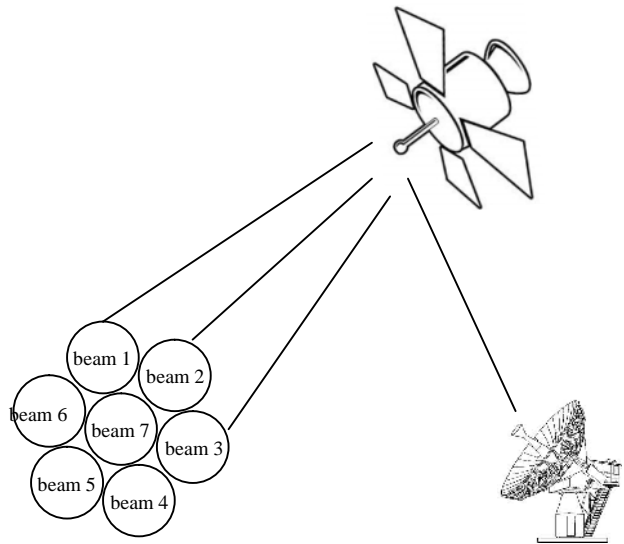

Figure 2. Multibeam satellite communication.

## 4. MULTI-USER DETECTION

The multiuser detection techniques developed are composed of the following operations: channel decoding, channel estimation, interference cancellation and signals combining. These iterative algorithms are formed of multistage architecture. In each stage, these operations are done successively.

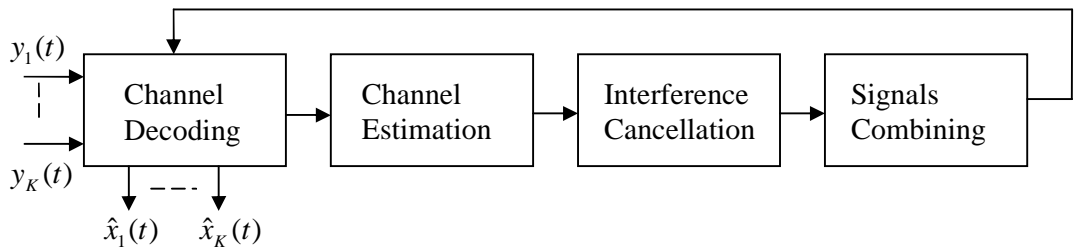

Figure 3. Receiver architecture.

### 4.1. Channel Decoding

We have implemented and matched two different decoding techniques for the convolutional channel code which are the Viterbi algorithm [29], [30], and the BCJR one [31], [32]. The second method needs SNR knowledge or in general the signal to noise plus interference ratio (SNIR).

#### 4.1.1. SNR estimation

To estimate the signal to noise ratios in the case of BCJR or MAP (Maximum a Posteriori) decoding algorithm, we have proposed a solution which is based on this algorithm itself. This

167

pilot aided method gives an estimation of the SNR or SNIR by minimizing the BER between the received and the training sequence signals. Then, the estimation of SNR is given by:

$$\hat{SNR} = \arg\left(\min_{snr \in [snr_{\min}\ snr_{\max}]} (BER)\right) \quad (12)$$

### 4.1.2. Phase compensation

The channel decoding algorithms process the quantity expressed in equation (11). Thus, phase compensation is needed. In the first stage of the multiuser detection techniques, we can use an estimation of the signals phases by processing the following pilot aided phase estimation. For the $k^{th}$ signal we have:

$$\hat{\varphi}_k = ang\left(\sum_{N_p} y_k(j) P_k(j)^*\right) \quad (13)$$

With, $P_k$ is the vector of $k^{th}$ user training sequence which contains $N_p$ pilot symbols. And $ang(.)$ denotes the angle of a complex number operator.

### 4.2. Channel Estimation

Channel estimation operation is to determine the coefficients of the channel matrix $H$. Then, the estimated coefficients will be used to cancel the MAI and compensate the channel effects. This technique works in an iterative process, where channel coefficients are updated at each iteration. Channel estimation of the $k^{th}$ signal at the $n^{th}$ iteration is given by:

$$\hat{h}_k^{(n)}(i) = \frac{1}{2 \times (N_i + N_p)} \sum_{j=1}^{N_i + N_p} y_k(j) \hat{x}_i^{(n)}(j) \quad (14)$$

Where,

- $\hat{h}_k^{(n)}(i)$ is the estimation of interference coefficient of signal i on signal k at $n^{th}$ iteration, or the (k,i)$^{th}$ element estimation of the matrix $H$ at the at the $n^{th}$ iteration.
- $N_i$ et $N_p$ are respectively the information symbols number and the training sequence number.
- The value 2 in the denominator is to compensate the effect on the channel coefficients estimation of the square of the QPSK symbols modulus.
- $\hat{x}_i^{(n)}(j)$ is the estimation of the $j^{th}$ symbol of user i at the $n^{th}$ iteration.

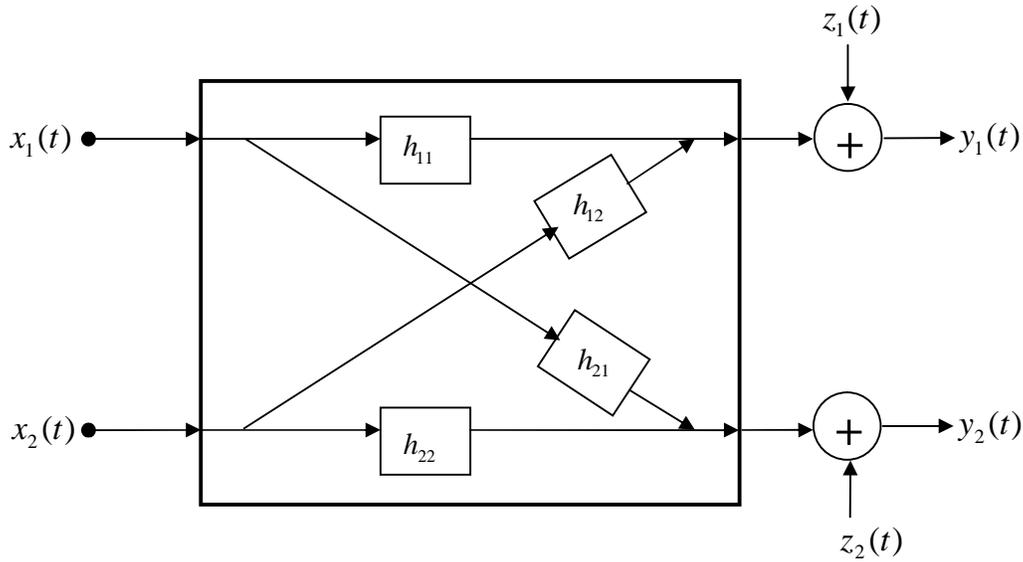

Figure 4. CCI Channel Model for K=2.

### 4.3. Interference Cancellation

In the interference cancellation bloc, users are dealt successively in each stage. Data of interfering signals are estimated and extracted from the considered signal with use the appropriate estimated channel coefficients.

Interference cancellation for $k^{th}$ signal at the $n^{th}$ iteration is developed in the following equation:

$$y_k^{(n)}(j) = y_k(j) - \sum_{k'=1}^{k-1} \hat{h}_k^{(n-1)}(k') x_{k'}^{(n)}(j) - \sum_{k'=k+1}^{K} \hat{h}_k^{(n-1)}(k') x_{k'}^{(n-1)}(j) \qquad (15)$$

### 4.4. Signals Combining

The signals combining operation ensures extraction of the useful signal parts which are interfered on the other beams, and adds them to this useful signal in order to improve the signal to noise plus interference ratio, which leads to detection improving. For the $j^{th}$ symbol of $k^{th}$ user at the $n^{th}$ iteration recombining expression can be given by:

$$R_k^{(n)}(j) = \sum_{\substack{k'=1 \\ k' \neq k}}^{K} \hat{h}_{k'}^{(n-1)}(k)^* \left( y_{k'}(j) - \sum_{\substack{k''=1 \\ k'' \neq k'}}^{K} \hat{h}_{k'}^{(n-1)}(k'') \hat{x}_{k''}^{(n-1)}(j) \right) \qquad (16)$$

Where, $(.)^*$ denotes the conjugate operator of complex number.
Thus, after recombining, the relative quantity of the $j^{th}$ symbol of $k^{th}$ signal, which will be dealt by the decoder at the $n^{th}$ iteration is expressed by :

$$\tilde{y}_k^{(n)}(j) = y_k^{(n)}(j) + R_k^{(n)}(j) \tag{17}$$

## 5. SIMULATIONS RESULTS

We evaluated the performance of the proposed solution by computer simulations. We considered a synchronous multibeam satellite communication scenario. The parameters of the simulation are inspired of the DVB RCS standard [33]. So we have used the convolutional coding employed by this system. This code is generated by two polynomials whose octal representations are: [171 133]. The modulation is QPSK. The other parameters used are: K=5, $N_i$=100 and $N_p$ =30. We have launched Monte Carlo simulations. In each iteration, we generated N= $N_i$ + $N_p$ symbols and K carrier phases uniformly distributed in $[0, 2\pi]$. The sequence of QPSK symbols is coded and interleaved before transmission. The interferences are supposed uniformly distributed between the co-frequency signals. Thus, the channel matrix can be presented by:

$$|H| = \begin{bmatrix} 1 & \varsigma & \varsigma & \varsigma & \varsigma \\ \varsigma & 1 & \varsigma & \varsigma & \varsigma \\ \varsigma & \varsigma & 1 & \varsigma & \varsigma \\ \varsigma & \varsigma & \varsigma & 1 & \varsigma \\ \varsigma & \varsigma & \varsigma & \varsigma & 1 \end{bmatrix}$$

Where, $\varsigma = 0.25$ and $a_1 = a_2 = \ldots = a_5 = 1$.

We have presented in figure 5 the simulations results of the multi-user detection technique with Viterbi channel decoding. The detection quality of the symbols is improved from a stage to another. This solution shows good performances although the low SNRs. In figure 6, we have shown the simulation results of the algorithm with BCRJ channel decoding. The BER values confirm the robustness of the detection method. And, we remark that this second solution which is base on MAP decoding gives better results compared to the first. This is due to the channel decoding techniques. However, in the computer simulation operation, the first solution which is based on Viterbi algorithm is faster. Thus, performances improvement due to MAP decoding introduced important computing and more complexity at the receiver. In addition to that, SNR estimation operation which is needed for BCJR decoding, procreates more processing time. If we compare now, the quality detection with and without signals combining technique in figures Fif1 and fig, it is clear that solution improved the performance of the receivers. It allowed somehow the decreasing of iterations number or stage of the algorithms by converging to the desired bit error rates.

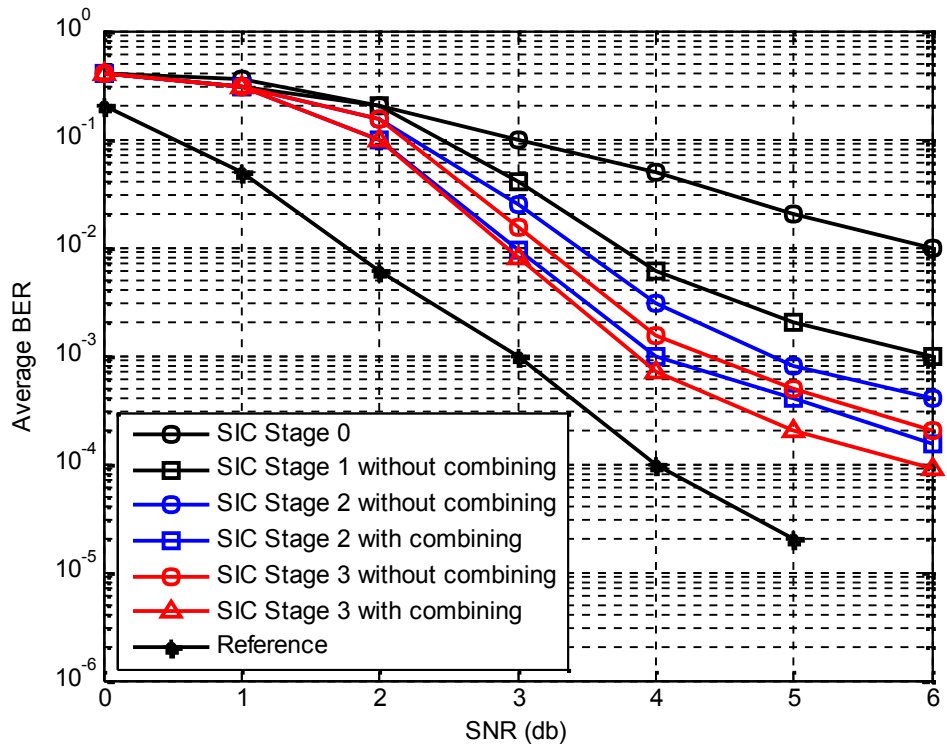

Figure 5. Performance evaluation of the SIC architecture with Viterbi decoding.

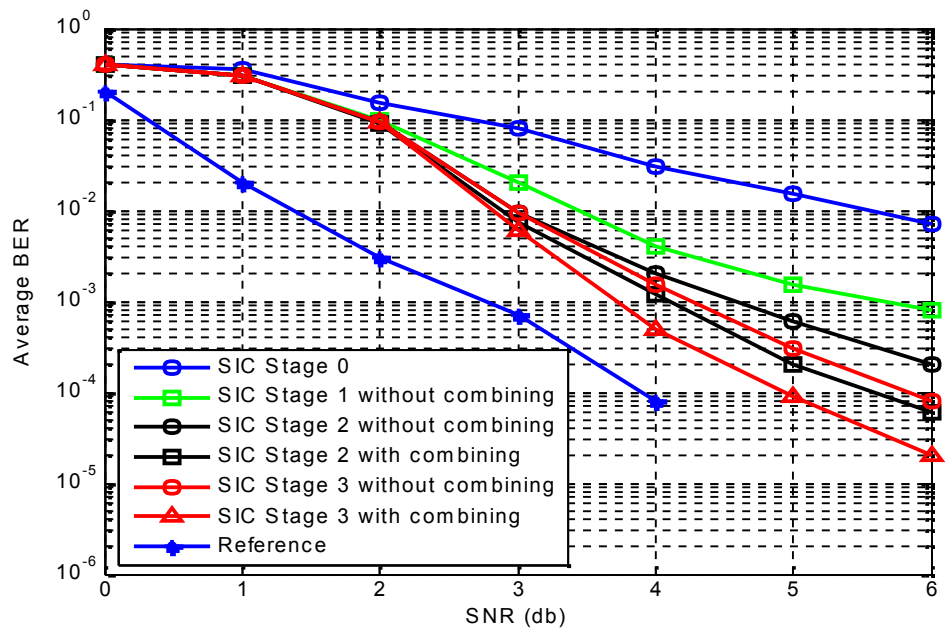

Figure 6. Performance evaluation of the SIC architecture with BCJR decoding.

## 6. CONCLUSIONS

In this paper, we have developed multiuser detection techniques for synchronous multibeam communications. These solutions are based on successive interference cancellation, channel estimation and convolutional decoding. Two channed decoding algorithms are implemented and matched. The viterbi algorithm and the BCJR one. Multiuser detection which adopts Viterbi decoding presented good performances in the presence of noise and MAI. The second solution, which is formed of BCJR algorithm as a channel decoding, showed its robustness against noise and CCI, and it gave low bit error rates, which are improved from a stage to another. Moreover, its performances are better than those of the method implementing Viterbi algorithm. But, this detection quality improvement is accompanied with more processing complexity. In fact, simulation results showed that the multiuser technique where Viterbi decoding is applied is faster than the method which contains MAP channel decoding. The simulation results presented also the importance of the signals combining technique. In fact, this operation improves the useful signal energy and, consequently the SNR increases which leads to better detection quality, independently of the convolution decoding technique incorporated. This signals combining solution benefits of spatial diversity due to antenna array. As a future work, we will concentrate on the case of asynchronous communications, and we can extend our work to parallel and hybrid architectures, and integrate eventually other channel estimation and decoding solutions.

## Authors

**Helmi CHAOUECH** received the engineering degree in telecommunications in 2006 and the M.S degree in communications systems in 2007 from the National Engineering School of Tunis, Tunisia. Currently, he is with the Innov'Com Laboratory at Higher School of Communications of Tunis (Sup'Com), Tunisia as a Ph.D student and he is an assistant in the Faculty of Economic Sciences and Management of Nabeul, Tunisia. His research interests include channel estimation, multiuser detection, wireless communication theory and multibeam satellite communications. 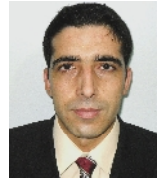

**Ridha BOUALLEGUE** received the Ph.D degree in electronic engineering from the National Engineering School of Tunis in 1998. In March 2003, he received the HDR degree in multiuser detection in wireless communications. From 2005 to 2008, he was the Director of the National Engineering School of Sousse, Tunisia. In 2006, he was a member of the national committee of science technology. Since 2005, he was the director of the 6'TEL Research Unit at Sup'Com.Since 2011, he was the founder and the director of Innov'Com Laboratory at Sup'com. Currently, he is the director of Higher School of Technology and Informatic, Tunis, Tunisia. His current rsearch interests include wireless and mobile communications, OFDM, space- time processing for wireless systems, multiuser detection, wireless multimedias communication and CDMA systems. 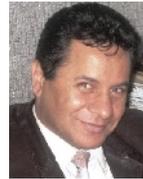